\documentclass[12pt]{article}
\usepackage{latexsym}
\begin{document}

\newtheorem{theorem}{\hspace{\parindent}Theorem}
\newtheorem{lemma}{\hspace{\parindent}Lemma}
\newtheorem{defin}{\hspace{\parindent}Definition}
\newtheorem{rem}{\hspace{\parindent}Remark}
\newtheorem{cor}{\hspace{\parindent}Corollary}
\newtheorem{prop}{\hspace{\parindent}Proposition}

\def\a{\alpha}
\def\b{\beta}
\def\g{\gamma}
\def\Ga{\Gamma}
\def\d{\delta}
\def\f{\phi}
\def\l{\lambda}
\def\th{\theta}
\def\p{\psi}
\def\pt{\xi}
\def\P{\Phi}   
\def\Th{\Theta}
\def\ph{\varphi}
\def\phat{\widehat{\p}}
\def\u{\widehat{u}}
\def\v{\widehat{v}}
\def\un{\widehat{u_n}}
\def\vn{\widehat{v_n}}
\def\y{{\bf Y}}
\def\r{{\bf R}}
\def\pu{\bar\pt}
\def\m{{\bf m}}
\def\e{{\bf e}}
\def\xxi{{\bf k}}
\def\x{{\bf x}}
\def\o{{\bf o}}
\def\r{{\bf r}}
\def\ss{{\bf s}}

\def\F{\mathcal{F}}
\def\G{\mathcal{G}}
\def\A{\mathcal{A}}
\def\X{\mathcal{X}}
\def\C{\mathcal{C}}
\def\M{\mathcal{M}}
\def\B{\mathcal{B}}
\def\H{\mathcal{H}}
\def\S{\mathcal{S}}
\def\T{\mathcal{T}}
\def\U{\mathcal{U}}

\def\pk{P_{k_0}}
\def\qk{Q_{k_0}}

\def\half{\frac{1}{2}}

\def\s{\sum_{k=2}^{\infty}}
\def\sk{\sum_{k=0}^{\infty}}
\def\sj{\sum_{j=-k}^{k}}
\def\sv{\sum_{m=-2}^{2}}
\def\sa{\sum_{\a=0}^{\infty}}
\def\sb{\sum_{\b=-\a}^{\a}}

\def\ds{\sigma(d\f)}
\def\ints{\int_{S^2}}
\def\lbo{\Delta_\m}
\def\grads{\nabla_\m}
\def\ddf{d\f^{n-1}}

\def\be{\begin{equation}}
\def\ee{\end{equation}}
\def\bea{\be\begin{array}{rcl}}
\def\eea{\end{array}\ee}
\def\beas{\begin{eqnarray*}}
\def\eeas{\end{eqnarray*}}

\def\df{\partial_\f}
\def\dt{\frac d{dt}}
\def\dx{\partial_{x_1}}
\def\dy{\partial_{x_2}}

\def\N{{\rm I\kern-.1567em N}}
\def\R{{\rm I\kern-.1567em R}}
\def\Z{{\rm Z\kern-.3em Z}}
\def\lsim{\mathrel{\rlap{\lower4pt\hbox{\hskip1pt$\sim$}}
    \raise1pt\hbox{$<$}}}               

\newcommand{\prf}{\noindent\hspace{\parindent}
     {\bf Proof :\ }}
\newcommand{\prfe}{\hspace*{\fill} $\Box$

\smallskip \noindent}

\def\n#1{\vert #1 \vert}
\def\nn#1{\left\Vert #1 \right\Vert}
\def\bn#1{\n{ #1 }_f}
\def\nbn#1{\nn{ #1 }_a}
\def\pn#1{\nn{ #1 }_{H^{-1/2}}}
\def\lnn#1{\nn{ #1 }_{L^2}}
\def\sn#1{\nn{ #1 }_{\infty}}
\def\hn#1{\nn{ #1 }_{H^s}}
\def\av#1{\left< #1 \right>_{\p}}

\title{Global Dissipativity and Inertial Manifolds for Diffusive Burgers Equations with Low-Wavenumber Instability}
\author{Jesenko Vukadinovic
\thanks {Department of Mathematics, CUNY-College of Staten Island, 1S-208, 2800 Victory Boulevard, Staten Island, NY 10314, USA. Tel: 718-982-3632.  Email: vukadino (at) math.csi.cuny.edu}}
\pagestyle{myheadings} \markboth{}{On diffusive Burgers equations with low wavenumber instability}
\maketitle

\begin{abstract}
Global well-posedness, existence of globally absorbing sets and existence of inertial manifolds is investigated for a class of diffusive Burgers equations.  The class includes diffusive Burgers equation with nontrivial forcing, the Burgers-Sivashinsky equation and the Quasi-Stedy equation of cellular flames. The global dissipativity is proven in 2D for periodic boundary conditions. For the proof of the existence of inertial manifolds, the spectral-gap condition, which Burgers-type equations do not satisfy in its original form is circumvented by the Cole-Hopf transform.  The procedure is valid in both one and two space dimensions. 
\end{abstract}
{{\em Mathematics subject classification:} 35K55, 35B41, 35B42,  37L25

\section{Introduction} 
We study a class of diffusive Burgers equations with low-wavenumber instability.  These are modified Burgers equations, which in addition to the usual Burgers nonlinearity and the diffusive term contain an additional linear term responsible for linearized instability of low wavenumbers.  The most prominent example of an equation of this type is the Kuramoto-Sivashinsky equation (KSE), 
\[
\partial_t U+(U\cdot\nabla)U+(\Delta^2+\alpha\Delta) U=\Delta U, 
\]
subject to appropriate initial and boundary conditions.  KSE has been introduced in \cite{S} as a model for  the flame front propagation in combustion theory.  Ever since,  the equation has been the subject of intense scientific interest  due to its complicated and interesting dynamical behavior, which for a range of parameters, is characterized by high-dimensional cellular chaos.  Many of the basic questions for the KSE in one space dimension subject to periodic boundary conditions, such as the global dissipativity and the existence of inertial manifolds have since been resolved (see e.g. \cite{CEES, G, FNST2}). In two space dimensions, the equation is not known to be globally well-posed.  Most of the recent research activity has shifted towards obtaining the best bounds on the absorbing sets and the dimensions of the global attractor and the inertial manifolds, since the numerical simulations suggest better estimates than the ones which have been obtained analytically. 

KSE is not the only equation of this type which is of interest.  The so-called Burgers-Sivashinsky equation (BSE),
\[
\partial_t U+(U\cdot\nabla)U=\Delta U+(\alpha-1)U, 
\]
was introduced as the simplest prototype of diffusive Burgers equations with low-wavenumber instability.  As it turns out, the similarity between KSE and BSE  due to the  instability of low wavenumbers, high wavenumber damping, and nonlinear stabilization via energy transfer from the low to the high modes  
is rather superficial.  The BSE is a gradient system, and as such it displays fairly simple dynamics characterized only by steady-states. This fact suggests that whatever the cause of cellular-chaotic dynamics may be, it is not simply due to the interplay between  diffusivity, Burgers energy transfer between Fourier modes, and low-wavenumber instability, but there is something more subtile at play.  It also suggests that estimates on the attractor obtained using methods which simultaneously apply to both KSE and BSE are not sharp, and certain mechanisms unique to KSE need to be identified and explored. 

In recent years, yet another system of the type described  above has become subject of interest.  The so-called Quasi-Steady equation (QSE), 
\[
\partial_t U+(U\cdot\nabla)U=\Delta U+\alpha(I-(I-\Delta)^{-1})U
\]
was introduced by Brauner et al. in \cite{BFHS} as the weakly nonlinear version of a certain truncation of the $\kappa-\theta$ model of cellular flames (see \cite{FGS}).  However, despite its ad-hoc introduction, it turns out that its dispersive relation 
\[
\omega(k)=-k^2+\frac{\alpha k^2}{1+k^2}=(\alpha-1)k^2-\alpha k^4+\mathcal{O}(k^6)+\dots,\ \ k\ll1,
\] 
most closely mimics the dispersion relation of the original combustion problem.  Just like KSE, QSE also models the onset of the cellular-chaotic dynamics; additionally however, it may better capture some important features of the global dynamics of the original combustion problem.   Unlike KSE and BSE, QSE is an integro-differential equation and it is nonlocal.  In some respect, it is more benign than the KSE; as we shall prove later in this paper, it is globally well posed and globally dissipative in two space dimensions, when subjected to periodic boundary conditions. 

Dissipative Burgers equations with low-wavenumber instability are yet another example of  parabolic PDEs, which exhibit long-term dynamics with properties typical of finite-dimensional dynamical systems. The global attractor, often considered  the central object in the study of long-term behavior of dynamical systems, appears to be inadequate in capturing this finite-dimensionality, even when its Hausdorff dimension is finite.  This is mainly due to two facts.  Firstly, the global attractor can be a very complicated set, not necessarily a manifold; the question whether  the dynamics on it can be described by a system of ODEs is yet to be resolved in the literature.   Secondly, although all solutions approach this set, they do so at arbitrary rates, algebraic or exponential, and, consequently, the dynamics outside the attractor is not tracked very well on the attractor itself.   When they exist, inertial manifolds emerge as most adequate objects to capture the finite-dimensionality of a dissipative parabolic PDE.  Introduced by Foias et al. in \cite{FST}, they are defined to remedy the  shortcomings of the global attractor just described: they should be finite dimensional positive-invariant Lipschitz manifolds which attract all solutions exponentially, and on which the solutions of the underlying PDE are  recoverable from solutions of a system of ODEs, termed `inertial form'.  The  existence of an inertial manifold does not merely have a theoretical value, but a practical one as well: using a system of ODEs instead of a system of PDEs facilitates computations and numerical analysis.  

Unfortunately, however useful, inertial manifolds seem unattainable for the vast majority of physically relevant parabolic PDE-s due to a rather restrictive spectral-gap condition that the system at hand is required to satisfy.  Despite the fact that this condition emerges naturally in any approach to the theory of inertial manifolds, the question of its actual importance remains unclear.  This can be demonstrated by comparing KSE and BSE.  The prior equation is known to satisfy the spectral-gap condition and therefore possess inertial manifolds despite its complicated attractor, while the second with its very simple dynamical behavior is not known to possess them.  One of the goals of this paper is to demonstrate that the spectral gap condition which the BSE does not satisfy is actually an artifact in this case.  The equation can be transformed into a form for which the condition will be satisfied.  Using the Cole-Hopf transform, a large class of diffusive Burgers equations including BSE and QSE will be shown to possess inertial manifolds in both one and two space dimensions. The idea to transform the equation was also developed and used by the author to prove the existence of inertial manifolds for a class of nonlinear Fokker-Planck equations, which appear in the modeling of nematic polymers (see \cite{V1, V2, V3}).  Rather than the Cole-Hopf transform, the author developed a nonlinear nonlocal transform which also eliminates the first-order derivatives from the equation, thus allowing the equation to satisfy the spectral-gap condition.  

Let us remark here that the inertial manifold result of this paper also applies to the diffusive Burgers equation with forcing. In the literature on inertial manifolds, this equation is often cited as an example of an equation with trivial dynamics, for which the existence of inertial manifolds is not known due to the spectral-gap condition.  Let  also be noted here  that the transformation approach is not an entirely new approach. In the paper \cite{Kwak2}, the author uses a different kind of transformation to prove the existence of inertial manifolds for the Burgers equation with forcing; the result, however, has since been shown to be flowed. 

The paper is structured as follows.  In Section 2, we introduce a more general framework for the study of diffusive Burgers equations with low-wavenumber instability subject to periodic boundary conditions; we shift our attention to a subclass of such equations, for which the low-wavenumber instability is due to a term $TU$, where $T:L^p\to L^p$ is a bounded Fourier multiplier for some $p>2$. Note that KSE is no longer in this class, and the approach presented in this paper does not provide an alternative proof of the existence of inertial manifolds for KSE.  In Section 3, expanding on the ideas of \cite{M}, we prove the global well-poedness and the global dissipativity for equations in this class in two space dimensions.  The procedure in one space dimension is simpler, and we omit it.  Finally, in Section 4, we transform the equation into a form which satisfies the spectral-gap condition, and prove the existence of inertial manifolds in both one and two space dimensions. 

\setcounter{equation}{0}
\section{Class of Diffusive Burgers Equations with Low-Wavenumber Instability}
\setcounter{equation}{0}
We examine global well-posedness, global dissipativity  and the existence of  inertial manifolds for a class of diffusive Burgers equations in one or two space dimensions, which assume the general form 
\be\label{burgers1}
\partial_t U+(U\cdot\nabla)U=\Delta U+TU+\nabla G, 
\ee
and where $U=\nabla \f$ for some $\f$. Recall that in two space dimensions, this is guaranteed if $\textrm{curl } U=0$, i.e., if $\dy u_1=\dx u_2$, and we write (\ref{burgers1}) as a system in the following form: 
\bea\label{burgers2}
\partial_t u_1+u_1\dx u_1+u_2\dy u_1&=&\Delta u_1+Tu_1+\dx G\\
\partial_t u_2+u_1\dx u_2+u_2\dy u_2&=&\Delta u_2+Tu_2+\dy G\\
\dy u_1=\dx u_2.
\eea
We consider the equation on  $Q=[-L/2,L/2]^d$ for some $L>0$ in space dimension $d=1$ or $d=2$, subject to periodic boundary conditions,
\[
U(t, \x+L\e_i)=U(t,\x), \ \ \ \ (t, \x)\in\R^+\times\R^d,\ \ \ \ i=1,2,\dots d,
\]
where $(\e_i)_{i=1}^d$ is the canonical basis of $\R^d$, and subject to the zero mean condition, 
\[
\int_Q U(t,\x)\ d\x=0.
\]

The term $TU$ is responsible for the instability of low wavenumbers.  The operator $T$ is assumed to be a Fourier multiplier associated with a given bounded symbol $m:\Z^d\to\R$ with zero mean ($m({\bf 0})=0$), and $G$ is a given $L$-periodic function with zero mean.  Due to the fact that the Fourier multipliers commute, the equation is often written in an integrated form 
\[
\f_t+\half \n{\nabla\f}^2=\Delta \f+T\f+G. 
\]
This form will be useful to us in proving the existence of inertial manifolds.  However, this equation does not preserve zero mean, and the  the dissipativity of (\ref{burgers1}) does not imply the dissipativity of this integrated form.  There are two possibilities to circumvent this difficulty. One is to use the following integrated form instead, which does preserve the mean: 
\be\label{burgers5}
\f_t+\half \n{\nabla \f}^2=\Delta \f+T\f+\half\int_Q \n{\nabla\f(\x)}^2\ d{\x}+G.  
\ee
Another approach, which we shall adopt in this paper, is to use the following integrated form, for which the dissipativity is implied by the dissipativity of (\ref{burgers1}): 
\be\label{burgers3}
\f_t+\langle\f\rangle+\half\n{\nabla \f}^2=\Delta \f+T\f+G.  
\ee
Here, $\langle\f\rangle=\frac1{L^d}\int_{Q} \f(\x)\ d\x$.  Observe that the mean then satisfies 
\be\label{mean}
\langle\f\rangle_t+\langle\f\rangle+\half\langle\n{\nabla \f}^2\rangle=0.  
\ee

As discussed in the introduction, there are several different equations of type (\ref{burgers1}), which are of interest.  The simplest example is the diffusive Burgers equation, i.e. the case when $T=0$.  This equation is dynamically very simple; the attractor consists of a single steady-state. However, even in this trivial case, the existence of an inertial manifold has yet not been established for a nontrivial forcing.  The attractor is not necessarily an inertial manifold, since it does not attract all solutions at an exponential rate.  As already discussed in the introduction, dynamically more interesting is the Burgers-Sivashinski equation, $m(\xxi)=(\alpha-1)$, $\alpha>1$, and much more interesting is the Quasi-Steady equation of cellular flames, 
\[
m(\xxi)=\frac{\alpha(2\pi/L)^2\n{\xxi}^2}{1+(2\pi/L)^2\n{\xxi}^2}.
\]
Both systems resemble the Kuramoto-Shivasinski equation, 
\[
m(\xxi)=\alpha (2\pi/L)^2\n{\xxi}^2\left(1 -(2\pi/L)^2\n{\xxi}^2\right).
\]
\section{Mathematical Setting and Existence of Absorbing Sets} 
For the study of the system (\ref{burgers1}), we develop the following mathematical fraimwork.  
For $s\geq0$, let us denote $\H^s=H_{\rm {per}}^s([-L/2,L/2]^d)$, and let $\nn{\cdot}_{\H^s}$ denote its norm. Let us use $H^s$ to denote the space 
\[
H^s=\left\{U:\R^d\to\R^d : U=\nabla \f \textrm{ for some }\f \in H^{s+1}\right\},
\]
endowed with the usual $\nn{\cdot}_{H^s}:=\nn{\cdot}_{\dot H^s}$ norm. 
In case $d=2$, let us also define 
\[
H_{x_1}^s=\left\{u\in H^s: \int_{-L/2}^{L/2} u(x_1,x_2)\ dx_1=0,\textrm{ for a.e. }x_2\in[-L/2, L/2] \right\}, 
\]
and, symmetrically, $H_{x_2}^s$. For simplicity,  the $L^p$, $p\geq1$ norm will be denoted $\n{\cdot}_{L^p}$, and the $L^2$ norm simply $\n{\cdot}$.  We also denote $\H=\H^0$ and $H=H^0$. 
 
The standard Galerkin prcedure can be used to establish local existence of solutions, and without proof we state the following 
\begin{lemma}
Suppose that  $T$ is the Fourier multiplier associated with a bounded symbol $m$, and $G\in H^1$.   For a given initial datum  $U_0=(u_{1_0},u_{2_0})\in H^1$ there exists a positive time $T=T(\n{\nabla U_0})$ such that (\ref{burgers1}) has a unique solution $U\in L^\infty(0,T; H^1)\cap L^2(0,T; H^2)$.  Moreover, if $[0, T_*)$ is the maximal interval of existence of $U$, then $U\in C([0,T_*);H^1)\cap C((0,T_*);H^\infty)$. If $T_*<\infty$, then $\lim\sup_{t\to T_*}\n{\nabla U(t)}=\infty$. The map $S:U_0\mapsto U(t)$ is continuous from $H^1$ into $C([0,T_*);H^1)$.
\end{lemma}

The global well-posedness of systems of type (\ref{burgers1}) and the existence of a globally absorbing ball  is a subtile matter due to instability of low wavenumbers. These questions have been resolved  for the  1D Kuramoto-Sivashinsky equation, 1D Burgers-Sivashinsky equation, 1D Quasi-Steady equation and, in space dimension two, for the 2D Burgers-Sivashinsky equation (see \cite{BFHR, CEES, G, M}).  Here, we state and prove a theorem on global well-posedness and dissipativity for the equations of the type (\ref{burgers1}) in two space dimensions.  The proof expands on the proof of \cite{M} for the 2D Burgers-Sivashinsky equation.   For the sake of simplicity, we shall not concern ourselves with the size of constants with respect to the periodicity $L$, and we shall assume that $G=0$. 

Similarly to other proofs of the dissipativity of the systems with instability of low wavenumbers, our proof is also based on the following Poincar\'e-type inequality, which we adopt from \cite{M}.  It follows from the inequality for the 1D case obtained in \cite{CEES} after integrating with respect to the second space variable.   
\begin{prop}\label{prop}
For any $\beta>0$ there exists a `gauge' function $\P\in H^{\infty}_{x_1}$ depending on $x_1$ only and a constant $\gamma>0$,  so that for every $v\in H^1_{x_1}$, the following inequality holds: 
\[
\beta \n{v}^2\leq \n{\dx v}^2+\int_Q v^2\P_{x_1}+\gamma\int_{-L/2}^{L/2}\left(\int_{-L/2}^{L/2}v\P_{x_1}\right)^2. 
\]
\end{prop}
We are now ready to state and prove the global well-posedness and dissipativity result. 
\begin{theorem}\label{main1}
Suppose that the symbol $m$ is such that $T$ is an $L^p$ multiplier for some $p>2$. Then the Cauchy problem for the equation (\ref{burgers1}) is globally well-posed for any initial datum $U_0=(u_{0_1},u_{0_2})\in H^1$ and there exists $\rho>0$, and a positive time $T=T(\n{\nabla U_0})$ such that  for all $t\geq T$ 
\be\label{diss1}
\nn{U(t)}_{H^1}< \rho. 
\ee
\end{theorem}

\prf As already indicated, the proof expands on the proof of \cite{M}, and we shall omit some details contained in that paper, and we shall emphasize the differences arising from replacing the term $U$ by the more general term, $TU$, where $T$ is an $L^p$ multiplier for some $p>2$.  Let $K_q:=\n{T}_{L^q\to L^q}$, whenever it is finite.  The proof consists of two parts.  Firstly, one obtains a priori estimates on the quantities $\n{(\dx u_1)^+}$ and $\n{(\dy u_2)^+}$, which suffice to prove the global well-posedness. Subsequently,  one imployes the Poincar\'e-type  inequality from Proposition \ref{prop} to prove the existence of a globally absorbing ball in $H$.  The existence of a globally absorbing ball in $H^1$ follows easily.   

Let $U=(u_1,u_2)$ be a solution of (\ref{burgers2}) to the Cauchy problem $U(0)=U_0$ for an initial datum $U_0=(u_{1_0}, u_{2_0})\in H^1$, and let us assume that for some $T_*<\infty$, the interval $[0,T_*)$ is the maximal interval of existence of that solution.  In particular, this means that $\limsup_{t\to T_*^-} \n{\nabla U(t)}=\infty$.  Multiplying ({\ref{burgers1}) by $U$, integrating by parts, and employing  H\"older inequality imply 
\beas
\half\dt \n{U}^2+\n{\nabla U}^2&=&(TU,U)+\half\int_Q(\mbox{div } U)\n{U}^2\\
&\leq& K_2\n{U}^2+\half\n{(\mbox{div } U)^+}\n{U}_{L^{4}}^2. 
\eeas
The Ladyzhenskaya inequality, 
\[
\n{f}_{L^{4}}^2\lsim \n{f}\n{\nabla f}, \ \ \ f\in H^1,
\]
and Young's inequality imply further that for some constant  $C>0$ large enough, we have 
\be\label{divu}
\dt \n{U}^2+\n{\nabla U}^2\leq C(1+\n{(\mbox{div } U)^+}^2)\n{U}^2.
\ee
On the other hand, multiplication by $-\Delta U$, integrating by parts, and using the fact that $\dx u_2=\dy u_1$ imply
\beas
\half\dt \n{\nabla U}^2+\n{\Delta U}^2&=&(T\nabla U,\nabla U)+((U\cdot\nabla U), \Delta U)\\
&=&(T\nabla U,\nabla U)+\half\int_Q(\mbox{div } U)\n{\nabla U}^2+\int_Q \partial_{x_k}u_i\partial_{x_i}u_j\partial_{x_k}u_j\\
&\leq&(T\nabla U,\nabla U)+\frac52\int_Q(\mbox{div } U)^+\n{\nabla U}^2\\
&\leq&K_2\n{\nabla U}^2+\frac52\n{(\mbox{div } U)^+}\n{\nabla U}_{L^{4}}^2.
\eeas
Similarly as before, we have 
\be\label{divvu}
\dt \n{\nabla U}^2+\n{\Delta U}^2\leq C(1+\n{(\mbox{div } U)^+}^2)\n{\nabla U}^2.
\ee
It appears crucial to obtain some  control over the term $\n{(\mbox{div } U)^+}$.  To this end, let us denote $w_1=(\partial_{x_1} u_1)^+$ and $w_2=(\partial_{x_2} u_2)^+$.  Multiplying the first equation of (\ref{burgers2}) by $-\dx(w_1)^{p-1}$, integrating by parts, and using the fact that $\dx u_2=\dy u_1$, one obtains 
\beas
\frac1p\dt\n{w_1}_{L^p}^p&\leq&\int_Q (Tw_1)w_1^{p-1}-\left(1-\frac1p\right)\n{w_1}_{L^{p+1}}^{p+1}-\int_Q(\dx u_2)^2w_1^{p-1}+\frac1p\int_Q w_2w_1^p\\
&\leq& K_p\n{w_1}_{L^p}^p-\left(1-\frac1p\right)\n{w_1}_{L^{p+1}}^{p+1}+\frac1p\int_Q w_2w_1^p
\eeas
Proceeding symmetrically with the second equation, then adding the two inequalities, and using Young's's inequality, one has 
\[
\dt(\n{w_1}_{L^p}^p+\n{w_2}_{L^p}^p)\leq p(\n{w_1}_{L^p}^p+\n{w_2}_{L^p}^p)-(p-2)(\n{w_1}_{L^{p+1}}^{p+1}+\n{w_2}_{L^{p+1}}^{p+1}).
\]
Denoting $\alpha_p:=\n{w_1}_{L^p}^p+\n{w_2}_{L^p}^p$, one obtains the inequality 
\be\label{alpha}
\dot\alpha_p(t)\leq p\alpha_p(t)-\frac{p-2}{(2L^2)^{1/p}}\alpha_p(t)^{(p+1)/p}, 
\ee
which further implies
\[
\sup_{t\in[t_0,T_*)}\alpha_p(t)^{1/p}\leq\max\left(\alpha_p(t_0)^{1/p},(2L^2)^{1/p}(p/(p-2))\right).
\]
In particular, this implies that $\n{(\mbox{div } U)^+}$ is uniformly bounded on $[0,T_*)$.  This fact together with (\ref{divu}) implies that $\n{U}$ is uniformly bounded on $[0,T_*)$.  This, in turn, together with (\ref{divvu}), integration by parts and Young's inequality implies that $\n{\nabla U}$ is uniformly bounded on $[0,T_*)$. We have arrived at a contradiction to the assumption that $\limsup_{t\to T_*^-} \n{\nabla U(t)}=\infty$, and the global well-posdness follows.  

Let us now assume that $T_*=+\infty$.  In order to prove the global dissipativity using (\ref{divu}) and (\ref{divvu}), one would like to obtain an a priori bound over $\n{(\mbox{div }U)^+}$ independent of $\n{(\mbox{div } U_0)^+}$.  One easily checks that 
\[
\tilde\alpha_p(t):=\max\left(2L^2(2p/(p-2))^pt^{-p}, 2L^2(2p/(p-2))^p\right)
\]
is an upper solution to inequality (\ref{alpha}), thus yielding the following inequality  
\[
\max(\n{w_1(t)}_{L^p},\n{w_2(t)}_{L^p})\leq\alpha_p(t)^{1/p}\leq \frac{2p}{p-2}(2L^2)^{1/p},\ \ t\geq1.
\]
As a consequence, there also exists a constant $c_p>0$  depending on $L$ and $p$ so that 
\be\label{np}
\max(\n{w_1(t)},\n{w_2(t)})\leq c_p,\ \ t\geq1.
\ee

The next step is to use the Poincar\'e inequality from Proposition \ref{prop} to circumvent the difficulties arising from the  instability of low wavenumbers.  To this end, let $\P\in H_{x_1}^\infty$ initially be any function depending on the variable $x_1$ only, and let $v_1=u_1-\P$.  Then, $v_1$ satisfies 
\[
\partial_t v_1-\Delta v_1-Tv_1+v_1\dx v_1+u_2\dx u_2+\P\dx v_1+v_1\P_{x_1}+\P\P_{x_1}-\P_{x_1x_1}-T\P=0.
\]
Multiplying by $v_1$, integrating over $Q$ and noticing 
\[
\int_Qu_2\dx u_2 v_1= \int_Qu_2\dy u_1 v_1=-\half\int_Q\dy u_2v_1^2
\]
and 
\[
\int_Q v_1^2 \dx v_1=0
\]
one arrives at
\[
\half\dt \n{v_1}^2\leq-\n{\nabla v_1}^2+K_2\n{v_1} +\half\int_Qv_1^2(\dy u_2)^+-\half\int_Q v_1^2\P_{x_1}+\int_Q v_1(-\P\P_{x_1}+\P_{x_1x_1}+T\P). 
\]
Using the Ladyzhenskaya inequality, one obtains  
\[
\int_Qv_1^2(\dy u_2)^+\leq\n{v_1}_{L^{4}}^2\n{(\dy u_2)^+}\leq C c_p\n{v_1}\n{\nabla v_1},\ \ t>1, 
\]
and Young's inequality implies  
\[
\dt\n{v_1}^2\leq-\n{\nabla v_1}^2 -\int_Q v_1^2\P_{x_1}+C_1\n{v_1}^2+C_2, 
\]
where the constant $C_1$ depends on $p$ and $L$, and the constant $C_2$ on $\P$.  According to Proposition \ref{prop},  we can chose $\P\in H^\infty_{x_1}$ depending on $x_1$ and $\gamma>0$ so that for any $v\in H^1_{x_1}$
\[
(C_1+1) \n{v}^2\leq \n{\dx v}^2+\int_Q v^2\P_{x_1}+\gamma\int_{-L/2}^{L/2}\left(\int_{-L/2}^{L/2}v\P_{x_1}\right)^2.
\]
Using the fact that $\int_{L/2}^{L/2}\n{\dx u_1}=2\int_{L/2}^{L/2}\n{(\dx u_1)^+}$, we obtain 
\beas
\dt\n{v_1}^2&\leq&-\n{v_1}^2 +\gamma\int_{-L/2}^{L/2}\left(\int_{-L/2}^{L/2}\dx u_1\P\right)^2+C_2\\
&\leq&-\n{v_1}^2 +\gamma\n{\P}_{L^\infty}^2\int_{-L/2}^{L/2}\left(\int_{-L/2}^{L/2}\n{\dx u_1}\right)^2+C_2\\
&\leq&-\n{v_1}^2 +4\gamma\n{\P}_{L^\infty}^2\int_{-L/2}^{L/2}\left(\int_{-L/2}^{L/2}\n{(\dx u_1)^+}\right)^2+C_2\\
&\leq&-\n{v_1}^2 +4\gamma\n{\P}_{L^\infty}^2L^2\n{(\dx u_1)^+}^2+C_2\\
&\leq&-\n{v_1}^2+\rho_1/3,
\eeas
for $t>1$, where $\rho_1=4\gamma\n{\P}_{L^\infty}^2L^2c_p^2+C_2$.  Proceeding symmetrically with the second equation of (\ref{burgers2}) and adding the two equations yields the existence of the time $T_0=T_0(\n{U_0})>0$ so that for all $t\geq T_0$, $\n{U(t)}<\rho_1$. From this fact, inequality (\ref{np}), and Young's inequality, the conclusion of the theorem  follows easily from inequality (\ref{divvu}).  \prfe
Let us make the following remarks: 
\begin{rem}
The multiplier $T$ associated with the symbol
\[
m(\xxi)=\frac{\alpha(2\pi/L)^2\n{\xxi}^2}{1+(2\pi/L)^2\n{\xxi}^2}
\] 
which appears in the Quasi-Steady equation is an $L^p$ multiplier for all $p>1$, and the conclusion of the theorem holds for the 2D Quasi-Steady equation. 
\end{rem}
\begin{rem}
The existence of a compact global absorbing set implies the existence of a global attractor, a backward and forward invarint set which attracts all solutions uniformly on bounded sets of initial data.  Moreover, the Hausdorff dimension of the global attractor is finite. 
\end{rem}

The above theorem implies the following well-posedness and dissipativity result for the integrated equation (\ref{burgers3}). 
\begin{cor}\label{cor}
Suppose that the symbol $m$ is such that $T$ is an $L^p$ multiplier for some $p>2$. Then the Cauchy problem for the equation (\ref{burgers3}) is globally well-posed for any initial datum $\phi_0\in \H^2$ and there exist constants $r>0$ and $r_\infty>0$, and a positive time $T=T(\nn{\phi_0}_{\H^2})$ such that for all  $t\geq T$ 
\[
\nn{\phi(t)}_{\H^2}<r
\]
and
\[
\n{\phi(t)}_{L^\infty}<r_\infty. 
\]
In particular, the set $B:=B_{\H^2}(r)$ is an absorbing set for the equation (\ref{burgers3}). 
\end{cor}
\prf The first inequality of the corollary follows immediately from Theorem \ref{main1} and the equation for the mean (\ref{mean}), using the Poincar\'e inequality.  The second one follows then from Agomon inequality in two space dimensions, 
\be\label{Agmon's}
\nn{f-\langle f\rangle}_{L^\infty}\lsim \n{f-\langle f\rangle}^{1/2}\n{\Delta f}^{1/2},\ \ \ \ \ f\in \H^2.  
\ee
\prfe
\section{Inertial Manifolds}
Let us first recall the definition of inertial manifolds and a theorem on their existence.  
Let us denote $A=-\Delta: \H^2\to \H$.  Following the standard procedure, from the Fourier modes, $e^{2\pi i\xxi\cdot \x/L}$, $\xxi\in\Z^d$, one constructs  the complete set of eigenfunctions $w_k$, $k=0,1,2,\dots$ of the operator  $A$ corresponding to the eigenvalues $\l_k$ which belong to the set 
\[
\{\Lambda_1, \Lambda_2, \cdots\}=\{(2\pi/L)^2\n{\xxi}^2: \xxi\in \Z^d\}, 
\]
where $(\Lambda_k)_{k=0}^\infty$ is an increasing sequence.  The application of existing theory of inertial manifolds is contingent on the existence of sufficiently large gaps in this set.  If $d=1$, then $\Lambda_k=(2\pi/L)^2k^2$, and the spectral gaps $\Lambda_{k+1}-\Lambda_k=(2\pi/L)^2(2k+1)$.  If $d=2$, it is known that $\Lambda_k\sim k$ and, after passing to a subsequence,  $\log k\lsim\Lambda_{k+1}-\Lambda_k$ (see \cite{Richards}). Therefore, there are arbitrarily large gaps in space dimensions one and two.   

Consider an evolution equation of the form
\be\label{abstract}
u_t+Au=N(u),
\ee
where $N: \H^{\alpha}\to \H^{\beta}$, $0\leq\alpha-\beta\leq 1$, is locally Lipschitz-continuous.  We define the projection operators 
\[
P_n u=\sum_{k=0}^n (u,w_k)_\H w_k, 
\]
and
\[
Q_n=I-P_n. 
\]
\begin{defin}
An inertial manifold $\M$  is a finite-dimensional Lipschitz manifold which is positively invariant, 
\[
S(t)\M\subset\M, \ t\geq0,
\]
and exponentially attracts all solutions uniformly on any bounded set $\U\subset \H^\a$ of initial data, 
\[
\rm{dist}_{\H^\a} (S(t)u_0,\M)\leq C_\U e^{-\mu t},\ u_0\in \U, \ t\geq0.
\]
The inertial manifold is said to be asymptotically complete, if for any solution $u(t)$ there exists $v_0\in\M$ such that
\[
\nn{u(t)-S(t)v_0}_{\H^\a}\to0, \ \ t\to\infty.
\]
\end{defin}

There are various methods  for proving the existence of inertial manifolds.  Almost of them, except for some very special cases, require some kind of Lipschitz continuity of the nonlinearity $N$ and make use of a very restrictive spectral gap property of the linear operator $A$.  In the approach we adopt here, these two conditions yield the strong squeezing property, which, in turn, yields the existence of an inertial manifold which is obtained as a graph of a Lipschitz mapping. We state the following   
\begin{theorem}\label{theory}
Suppose that the nonlinearity $N$ in (\ref{abstract}) satisfies the following three conditions
\begin{itemize}
\item It has compact support in $\H^\a$, i.e. 
\[
\rm{supp}(N)\subset B_{\H^\alpha}(\rho)
\]
\item It is bounded, i.e. $\nn{N(u)}_{\H^\b}\leq C_0$ for $u\in\H^\a$
\item It is globally Lipschitz continuous, i.e. 
\[
\nn{N(u_1)-N(u_2)}_{\H^\b}\leq C\nn{u_1-u_2}_{\H^\a}
\]
 for $u_1, u_2\in\H^\a$. 
\end{itemize}
Suppose that the eigenvalues of $A$ satisfy the spectral gap condition, 
\be\label{sgc}
\l_{n+1}-\l_n>2C\left(\l_n^{\frac{\a-\b}2}+\l_{n+1}^{\frac{\a-\b}2}\right), 
\ee
for some $n\in \N$.  Then there exists an asymptotically complete  inertial manifold which is obtained as the graph of a Lipschitz function $\Phi:P_n H\to Q_n H$, i.e.
\[
\M=\G[\Phi]=\{p+\Phi(p) : p\in P_n \H\}
\]
with 
\[
\nn{\Phi(p_1)-\Phi(p_2)}_{\H^\a}\leq l\nn{p_1-p_2}_{\H^\a}. 
\]
Restricting (\ref{abstract}) to $\M$ yields the ordinary differential equation for $p=P_n u$
\[
\frac{dp}{dt}+Ap=P_n N(p+\Phi(p))
\]
termed the inertial form.  
\end{theorem} 
\prf As already mentioned, there are several different proofs of some version of this theorem.  For an elegant improvement of the geometric approach developed in \cite{CFNT2} see also \cite{Rob, Rob1}.\prfe

The above theorem does not apply to Burgers equation in its original form.  To illustrate this fact, let us proceed in the usual fashion and denote $N(U)=-(U\cdot\nabla)U+TU+\nabla G$.  The first three conditions of the theorem could be circumvented in the usual fashion by `preparing' the equation; instead of $N$, one checks the conditions of the theorem for $N_P$ which is obtained from $N$ in a manner which leaves it unchanged on a bounded absorbing set; it is declared equal to zero outside a larger ball which contains the absorbing set, and in between, it is defined in a fashion which would insure the global Lipschitz continuity.  It has been proven that this procedure does not change the dynamics inside the absorbing ball which all solutions enter exponentially fast. In order to be able to prepare the equation, on can easily see that we would have to choose $\a=1$ and $\b=0$. However, it can be easily checked that the spectral gap condition $(\ref{sgc})$ is not satisfied for a large Lipschitz constant $C$ in either space dimension $d=1$ or $d=2$.  

The main idea of this section is to circumvent the difficulty presented by the spectral-gap condition by performing the change of variables, 
\[
\Psi:\phi\mapsto\p=e^{-\half\phi},
\]
in (\ref{burgers3}) to obtain the equation  
\be\label{burgers5}
\p_t=\Delta \p+\p T[\log(\p)]-\p\int_Q \log(\p({\bf{y}}))\ d{\bf{y}}-\half\p G.  
\ee
This is the Cole-Hopf transform which eliminates the Burgers nonlinearity from the equation.  In case of the non-forced diffusive Burgers equation, it transforms it into a linear heat equation.  In this case, however, the result of the transform is a new nonlinear parabolic equation. Let us denote $\P=\Psi^{-1}$. 
 \begin{lemma}
Let $B=B_{\H^2}(r)\subset B_{L^\infty}(r_\infty)$ and $\B=\Psi(B)$, where $r$ and $r_\infty$ were introduced in Corollary \ref{cor}.  Then $\B\subset\H^2$ is a bounded set and $\Psi:(B, \nn{\cdot}_{\H^1})\to(\B, \nn{\cdot}_{\H^1})$ is a Lipschitz homeomorphism. 
\end{lemma}
\prf
Firstly, observe that  $r_0:=e^{-r_\infty/2}\leq\psi\leq e^{r_\infty/2}=:r_1$ holds point-wise for all $\psi\in\B$. Secondly, for $\p=e^{-\f/2}\in\B$, in view of the Ladyzhenskaya inequality, 
\[
\n{\Delta \p}=\n{(-\half\Delta\f+\frac14\n{\nabla\f}^2)e^{-\f/2}}
\leq \half r_1(\n{\Delta\f}+\n{\nabla\f}_{L^4}^2)
\leq Cr_1(\n{\Delta\f}+\n{\nabla\f}\n{\Delta\f}), 
\]
and we easily conclude that there exists $r_2>0$ so that $\B\subset B_{\H^2}(r_2)$. 

To see that the mapping $\Psi$ is a Lipschitz homeomorphism, let $\p_1=e^{-\f_1/2}\in\B$ and $\p_2=e^{-\f_2/2}\in\B$.  We then have 
\beas
\n{\nabla \p_1-\nabla \p_2}&=&\half\n{\nabla\f_1\p_1-\nabla\f_2\p_2}\\
&\leq&\half\n{(\nabla\f_1-\nabla\f_2)\p_1}+\half\n{\nabla\f_2(\p_1-\p_2)}\\
&\leq&\half r_1\n{\nabla\f_1-\nabla\f_2}+\half\n{\nabla\f_2}_{L^4}\n{\p_1-\p_2}_{L^4}\\
&\leq&\half r_1\n{\nabla\f_1-\nabla\f_2}+\frac14 r_1\n{\nabla\f_2}_{L^4}\n{\f_1-\f_2}_{L^4}\\
&\leq&C \nn{\f_1-\f_2}_{\H^2}. 
\eeas
In the last step we again used the Ladyzhenshaya and Young's inequalities, and the Lipschitz continuity of $\Psi|_B$ follows.  On the other hand,  
\beas
\n{\nabla \f_1-\nabla \f_2}&=&2\left|\frac{\nabla\p_1}{\p_1}-\frac{\nabla\p_2}{\p_2}\right|\\
&\leq&\frac2{r_0}\n{\nabla\p_1-\nabla\p_2}+\frac2{r_0^2}\n{\nabla\p_2(\p_1-\p_2)}\\
&\leq&\frac2{r_0}\n{\nabla\p_1-\nabla\p_2}+\frac2{r_0^2}\n{\nabla\p_2}_{L^4}\n{\p_1-\p_2}_{L^4}\\ 
&\leq&C \nn{\p_1-\p_2}_{\H^2}, 
\eeas
and the Lipschitz-continuity of $\P|_\B$ follows. 
\prfe
From Corollary \ref{cor} follows the following 
\begin{cor}\label{thm}
Suppose that the symbol $m$ is bounded and that $T$ is an $L^p$ multiplier for some $p>2$. Then the Cauchy problem for the equation (\ref{burgers5}) is globally well-posed for any positive initial datum  $\p_0\in \H^2$, and there exists a positive time $T=T(\nn{\log( \p_0)}_{\H^2})$ such that for all  $t\geq T$ we have  
$
\p(t)\in\B. 
$
\end{cor}

Define now 
\[
N(\p):=-\half\p T[\P(\p)]+\half\p\int_Q \P(\p({\bf{y}}))\ d{\bf{y}}-\half\p G.
\]
In view od the fact that that $\P:\B\to \H^1$ is a Lipschitz continuous, it is easy to check that $N|_\B:\B\to \H^1$ is Lipschitz continuous with respect to norm $\nn{\cdot}_{\H^1}$. We are now in the position to prepare the equation as it is customary in the literature. First, we modify the nonlinear term: 
\[
N_P(\p)=\left\{\begin{array}{ll}N(\p), & \mbox{if } \p\in \B\\
                     0, & \mbox{if } \p\in \H^1\backslash B_{\H^1}(2r_2)\end{array}\right.
\]
$N_P:\B\cup(\H^1\backslash B_{\H^1}(2r_2))\to\H^1$ is clearly a Lipschitz continuous. Denote by $C>0$ its Lipschitz constant.  Following \cite{WW}, a Lipschitz-continuous function defined on a subset of a Hilbert space can be extended to a Lipschitz continuous function defined on the entire Hilbert space, even preserving the Lipschitz constant $C>0$.  Without changing the notation, let us by $N_P:\H^1\to \H^1$ denote such an extension.  The prepared equation reads now
\be\label{pe}
\dt \p+A\p=N_P(\p).
\ee
This equation clearly satisfies the first three conditions of the Theorem \ref{theory}, where we chose $\alpha=\beta=1$.  The spectral gap condition (\ref{sgc}) reads in this case $\lambda_{n+1}-\lambda_{n}\geq 4C$, which is clearly satisfied for some large enough $n\in\N$ in both space dimensions one and two, as there exist arbitrary large spectral gaps of $A$ in both of these dimensions.  Thus, the prepared equation (\ref{pe}) possesses an asymptotically complete inertial manifold $\M_P$.  Following the general procedure of \cite{FNST2}, from the existence of an inertial manifold $\M_p$ for the equation (\ref{pe}) one infers the existence of an asymptotically complete inertial manifold $\M$ for the transformed equation (\ref{burgers5}).  Finally, since $\Psi$ is a Lipschitz-homeomorphism in $\H^1$, an asymptotically complete inertial manifold in $H^1$ for the integrated equation (\ref{burgers3}) is obtained as $\P(\M\cap\B)$. 
Finally,
\[
M=\nabla\M=\left\{\nabla\f:\f\in \P(\M\cap\B)\right\}
\]
is an asymptotically complete inertial manifold in $H$ for the equation (\ref{burgers1}). Thus, we have proved the following 
\begin{theorem}
Suppose that the symbol $m$ is such that $T$ is an $L^p$ multiplier for some $p>2$, and $G\in H^2$. Then the system (\ref{burgers1}) possesses an asymptotically complete inertial manifold.  
\end{theorem}

\section*{Acknowledgments}  
I would like to thank Igor Kukavica for inspiring discussions.  This work was supported in part by the NSF grant  DMS-0733126. 


\begin{thebibliography}{99}

\bibitem{BFHR}
C.-M. Brauner, M. Frankel, J. Hulshof and V. Roytburd, {\it Stability and attractors for the quasi-steady equation of cellular flames}, ``Interfaces and Free Boundaries", {\bf 8} (2006), 301--316

\bibitem{BFHS}
C.-M. Brauner, M. Frankel, J. Hulshof and G. I. Sivashinsky, {\it Weakly nonlinear asymptotics of the $\kappa-\theta$ model of cellular flames: the QS equation}, ``Interfaces and Free Boundaries" {\bf 7} (2005), 131--146

\bibitem{CLS}
S. N. Chow, K. Lu and G. R. Sell, {\it  Smoothness of inertial manifolds}, ``J.
Math. Anal. Appl.'', {\bf 169} (1992), 283--312

\bibitem{CEES} 
P. Collet, J. P. Eckmann, H. Epstein and J. Stubbe, {\it Globally attracting set for the Kuramoto-Sivashinsky equation}, ``Commun. Math. Phys.", {\bf 152} (1993), 203--214

\bibitem{Const1}
P. Constantin, {\it A construction of inertial manifolds}, ``Contemporary
Mathematics", {\bf 99} (1989),  27--62


\bibitem{CF}
P. Constantin and C. Foias, ``Navier-Stokes Equations", 
University of Chicago Press: Lectures in Mathematics, Chicago/London, 1988. 


\bibitem{CFNT2}
P. Constantin, C. Foias, B. Nicolaenko and R. Temam ``Integral and
inertial manifolds for dissipative partial differential equations", 
Springer-Verlag, Applied Math. Sciences {\bf 70},  New York


\bibitem{CFNT1}
P. Constantin, C. Foias, B. Nicolaenko and R. Temam, {\it Spectral
barriers and inertial manifolds for dissipative partial
differential equations},  ``J. Dynam Diff. Eq.'',  {\bf 1} (1988), 45--73



\bibitem{FJKT}
C. Foias, M. S. Jolly, I. Kukavica and E. S. Titi, {\it The Lorenz
equation as a methafor for some analytic, geometric, and
statistical properties of the Navier-Stokes equations},  
``Discrete Cont. Dyn. Syst.",  {\bf 7}(2) (2001), 403--30

\bibitem{FNST2}
C. Foias, B. Nikolaenko, G. R. Sell and R. Temam, {\it Inertial manifolds for
the Kuramoto-Sivashinsky equation and an estimate of their lowest dimension}, ``J. Math. Pures Appl.'',  {\bf 67} (1988), 197--226


\bibitem{FST}
C. Foias, G R. Sell and R. Temam, {\it Vari\'et\'es inertielles des \'equations differe\'entielles dissipatives},  ``C. R. Acad. Sci. Paris I", {\bf 301} (1985), 285--288



\bibitem{FST1}
C. Foias, G R. Sell and R. Temam, {\it Inertial manifolds for
nonlinear evolutionary equations}, ``J. Diff. Eq.", {\bf 73} (1988) 309--53

\bibitem{FGS}
M. Frankel, P. V. Gordon and G. I. Sivashinsky, {\it On disintegration of near-limit cellular flames}, ``Phys. Lett. A",  {\bf 310} (2003),  389--392

\bibitem{G}
J. Goodman, {\it Stability of the Kuramoto-Sivashinsky and related systems},  ``Commun. Pure Appl. Math.", {\bf 47} (3) (1994), 293-306

\bibitem{K1}
I. Kukavica,  {\it On the behavior of the solutions of the
Kuramoto-Sivashinsky equations for negative time}, ``J. Math.
Anal. Appl.", {\bf 166} (1992) 601--06

\bibitem{K2}
I. Kukavica,  {\it On Fourier parameterization of global attractors for equations in one space dimension},  ``Discrete Cont. Dyn. Syst.",  {\bf 13} (3) (2005), 553--560


\bibitem{Kwak1} 
M. Kwak, {\it Finite dimensional inertial forms for the 2D Navier-Stokes Equations}, ``AHPCRC Preprint 91-30, Minnesota", (1991)

\bibitem{Kwak2} 
M. Kwak,  {\it Finite dimensional description of convective reaction-diffusion equations}, ``J. Dynam. Differential Equations ", {\bf 3} (3) (1992), 515--543


\bibitem{MPS}
J. Mallet-Paret and G. R. Sell, {\it Inertial manifolds for reaction diffusion
equations in higher space dimensions}, ``J. Amer. Math. Soc.", {\bf 1}  (1988), 805--866

\bibitem{MPSS} 
J. Mallet-Paret, G. R. Sell and Z. Shao, {\it Obstructions to the existence of normally hyperbolic inertial manifolds}, ``AHPCRC Preprint 93-041, Minnesota" (1993)

\bibitem{M} 
L. Molinet, {\it A bounded global absorbing set for the Burgers-Sivashinsky equation in space dimension two},  ``C.R. Acad. Sci. Paris I", {\bf 330} (2000),  635--640

\bibitem{Richards} 
I. Richards, {\it On the gaps between numbers which are the sum of two squares}, ``Adv. Math.",  {\bf 46} (1982), 1--2

\bibitem{Rob}
J. C. Robinson,  {\it Inertial manifolds and the cone condition}, ``Dyn. Systems Appl.", {\bf 2} (1993) 311--30.

\bibitem{Rob1}
J C. Robinson, {\it A concise proof of the ÒgeometricÓ construction of inertial
manifolds},  ``Phys. Lett. A",  {\bf 200} (1995), 415--17

\bibitem{S}
G. I. Sivashinsky, {\it Nonlinear analysis of hydrodynamics instability in laminar flames Part I. Derivation of basic equations}, ``Acta Astronaut.", {\bf 4} (1997), 1177--1206 
 
\bibitem{Temam1} 
R. Temam, ``Infinite dimensional dynamical systems in Mechanics and Physics", Springer-Verlag, Applied Math. Sciences {\bf 68}, New York, 1988

\bibitem{Temam2}
R. Temam, {\it Inertial manifolds},  ``Math. Intell.", {\bf 12} (1990), 68--73


\bibitem{V1} 
J. Vukadinovic {\it Finite-dimensional description of the long-term dynamics for the Doi-Hess model for rodlike nematic polymers in shear flows}, ``Commun. Math. Sci.",  {\bf 6} (4) (2008), 975--993

\bibitem{V2}
J. Vukadinovic, {\it Inertial Manifolds for a Smoluchowski Equation on the Unit Sphere}, ``Commun. Math. Phys.", {\bf 285} (2009), 975--990

\bibitem{V3}
J. Vukadinovic {\it Inertial manifolds for a Smoluchowski equation on a circle}, ``Nonlinearity" {\bf 21} (2008), 1533--1545


\bibitem{WW}
J. H. Wells and L. R. Williams, ``Embeddings and extensions in analysis. Ergebnisse der Mathematik und ihrer Grenzgebiete", Springer Verlag, New York-Hedelberg, 1975. 
\end{thebibliography}
\end{document}